\documentclass[aps, prd, 12pt, eqsecnum, tightenlines, notitlepage, nofootinbib, floatfix]{revtex4-1}

\usepackage[bookmarksnumbered=true,bookmarksopen=true]{hyperref}

\usepackage{multirow, slashed, xcolor, amsmath}

\usepackage{graphicx}
\graphicspath{{plots/}}

\definecolor{red}{rgb}{1,0,0}

\begin{document}

\title{Impact of transforming to Conformal Fermi Coordinates on Quasi-Single Field Non-Gaussianity}

\author{Adriano Testa and Mark B. Wise}
\affiliation{Walter Burke Institute for Theoretical Physics, California Institute of Technology, Pasadena, CA 91125}

\begin{abstract}
In general relativity predictions for observable quantities can be expressed in a coordinate independent way. Nonetheless it may be inconvenient to do so. Using a particular frame may be the easiest way to connect theoretical predictions to measurable quantities. For the cosmological curvature bispectrum such frame is described by the Conformal Fermi Coordinates. In single field inflation it was shown that going to this frame cancels the squeezed limit of the density perturbation bispectrum calculated in Global Coordinates. We explore this issue in quasi single field inflation when the curvaton mass and the curvaton-inflaton mixing are small. In this case, the contribution to the bispectrum from the coordinate transformation to Conformal Fermi Coordinates is of the same order as that
from the inflaton-curvaton interaction term but does not cancel it.
\end{abstract}
\maketitle

\section{Introduction}
In standard single field inflationary cosmology~\cite{starobinskii_spectrum_1979,guth_inflationary_1981,linde_new_1982,linde_coleman-weinberg_1982,albrecht_cosmology_1982} the cosmological density perturbations are almost Gaussian~\cite{maldacena_non-gaussian_2003}. Non-Gaussianities express themselves as connected parts of curvature perturbation correlation functions. The Fourier transform of the three point function of the curvature fluctuations is called\footnote{Up to a factor of $(2\pi)^3 \delta({\bf k}_1+{\bf k}_2+{\bf k}_3)$.} the bispectrum and is denoted by $B_\zeta({\bf k}_1,{\bf k}_2, {\bf k}_3)$. The bispectrum in standard single field inflation  was first calculated by Maldacena~\cite{maldacena_non-gaussian_2003} in Global Coordinates (GC) and it is suppressed by slow roll parameters.

A phenomenologically relevant limit of the bispectrum is the squeezed limit in which one of the wave-vectors ${\bf q}\equiv {\bf k}_1$ is very small in magnitude compared to the other two, $|{\bf q}| \ll |{\bf k}_{2,3}|$. Since ${\bf k}_1+{\bf k}_2+{\bf k}_3=0$ we have that ${\bf k}\equiv{\bf k}_2\simeq -{\bf k}_3$.
The squeezed limit of the bispectrum influences the galaxy power spectrum at small wavevectors \cite{dalal_imprints_2008}.

It has been shown~\cite{tanaka_dominance_2011,pajer_observed_2013, cabass_how_2017,bravo_vanishing_2018} that in standard single field inflation transforming to Conformal Fermi Coordinates (CFC)~\cite{pajer_observed_2013,dai_conformal_2015} with respect to the very long wave-length (small wave-vector) curvature perturbations cancels the squeezed limit of the bispectrum calculated in GC.
This cancellation is manifest in the de Sitter era before reheating takes place.
Many inflationary models have been studied that can give rise to significant non-Gaussianities, see for example~\cite{allen_non-gaussian_1987,bartolo_non-gaussianity_2002,alishahiha_dbi_2004,chen_large_2010,chen_quasi-single_2010,baumann_signature_2012,arkani-hamed_cosmological_2015,kumar_heavy-lifting_2018,deutsch_influence_2018,chen_neutrino_2018,an_sitter_2019,mcaneny_new_2019,kumar_cosmological_2019,lu_cosmological_2020, hook_minimal_2020}. One of the most studied and simplest of these is called Quasi Single Field Inflation (QSFI). It has an additional scalar field called the curvaton that mixes with the inflaton creating a rich dynamics that can lead to measurable curvature non-Gaussianities. We work in the limit where the mass of the curvaton and the coupling between the curvaton and the inflaton are small compared to the Hubble constant during inflation. We calculate the bispectrum in this limit in GC and then transform it to CFC.
The contribution from transforming to CFC and from the interaction vertex in GC are typically of the same order but do not cancel against each other\footnote{We expect the pure gravity contribution to be smaller.}.
Although QSFI can have large measurable non-Gaussianities, in the limit we work (where the potential interactions of the curvaton are negligible) $f_{\rm NL}$ is only about $10^{-2}$. Throughout this work, we use $G = c = 1$ units and $\eta_{\mu\nu}=\operatorname{diag}(-1, +1, +1, +1)$.
\section{Scale Invariance}
In this paper we consider a de Sitter background metric
\begin{equation}
    ds^2= -dt^2+e^{2Ht}dx^idx^i\,
\end{equation}
and we work in the limit where the Hubble constant during inflation ($H$) and the derivative with respect to the time $t$ of the inflaton field ($\dot{\phi}_0$) do not depend on time.
Expressing the metric in terms of the conformal time $\tau=-e^{-H t}/H$, we have
\begin{equation}
ds^2= \frac{1}{H^2\tau^2}\left( -d\tau^2+dx^idx^i \right)
\end{equation}
where the beginning and the end of inflation correspond to, respectively, $\tau \rightarrow -\infty$ and $\bar\tau\simeq 0$.
The background metric exhibits scale invariance under the transformation $\tau \rightarrow \lambda \tau$ and $x^i \rightarrow \lambda x^i$ that is preserved when $\dot \phi_0$ and $H$ are constant.
This symmetry implies that the power spectrum is a homogeneous function of order minus three in $1/\tau$ and $|{\bf p}|$. That is,
\begin{equation}\label{eq:scaleinvaraince}
\left(3+\frac{\partial}{\partial{\log{|{\bf p}|}}} \right) P_\zeta(\tau, |{\bf p}|) =\frac{\partial}{\partial{\log{\tau}}} P_\zeta(\tau, |{\bf p}|)\,.
\end{equation}
In this paper we will neglect the time evolution of $H$ and ${\dot \phi}_0$ that is important towards the end of inflation and depends on the shape of the inflaton potential. Hence all our results will be scale invariant. Scale invariance has implications for the higher point correlations of the curvature fluctuations as well. For example, it implies that the bispectrum $B_\zeta(\tau,{\bf k}_1,{\bf k}_2,{\bf k}_3)$ is a homogeneous function of ${\bf k}_1,{\bf k}_2,{\bf k}_3$ and $1/\tau $ of degree minus six.
\section{Quasi Single Field Inflation}\label{sec:QSFI}
In QSFI the inflaton field $\phi$ is accompanied by another scalar field the curvaton $s$. Although $s$ does not participate in the slow roll process, it does interact and mix with the inflaton through the term \cite{assassi_planck-suppressed_2014,an_non-gaussian_2018}
\begin{equation}
{\cal L}_{{\rm dim}~5}=- \frac{1}{\Lambda}{g^{\mu \nu}\partial_{\mu}\phi \partial_{\nu} \phi} s \,.
\end{equation}
We work in the gauge where the inflaton field is only a function of time $\phi_0(t)$ with no fluctuations. The Goldstone field $\pi(x)$, associated with time translational invariance breaking (by the time dependence of $\phi_0$)~\cite{cheung_effective_2008}\footnote{In~\cite{cheung_effective_2008} it is denoted by $\pi_c$.} gives rise to the curvature fluctuations $\zeta$ which are linearly related to $\pi$ via
 \begin{equation}
\zeta =-{\frac{H} {\dot \phi_0}}\pi \,.
 \end{equation}
In a de Sitter background, the Lagrangian describing $\pi(x)$ and $s(x)$ is then
\begin{align}
\mathcal{L} = \mathcal{L}_{0} + \mathcal{L}_{{\rm int}}\,
\end{align}
where
\begin{equation}
\label{fft}
\mathcal{L}_{0}=\frac{1}{2(H\tau)^{2}}\left[\left(\partial_{\tau}\pi\right)^{2} - \nabla\pi\cdot\nabla\pi + (\partial_{\tau}s)^{2} - \frac{m^{2}}{(H\tau)^{2}}s^{2}-\nabla s\cdot\nabla s-\frac{2\mu}{H\tau}s\partial_{\tau}\pi\right]\,
\end{equation}
and
\begin{equation}
\label{intlagrange}
{\cal L}_{\rm int}  =  \frac{1}{\Lambda(H\tau)^{2}} \left[(\partial_{\tau}\pi)^{2}- {\nabla { \pi}} \cdot  {\nabla \pi} \right]s\,.
\end{equation}
Note that we have neglected any potential interaction terms for the curvaton $s$. In Eq.~\eqref{fft} we introduced
\begin{equation}\label{eq:mu}
 \mu=2{\dot { \phi}_0}/{{ \Lambda}}
 \end{equation}
and we rescaled $\pi$ by $\dot{\phi}_{0}$ (we take $\dot{\phi_0} >0$) to obtain a more standard normalization for the $\pi$ kinetic term.  We have also included the measure factor $\sqrt{-g}$ in the Lagrangian so that the action is equal to $\int d^3x  d\tau  {\cal L}$.
The  kinetic mixing term between $\pi$ and $s$ in Eq.~\eqref{fft} is the result of the background inflaton field breaking Lorentz invariance.
We now introduce the quantities
\begin{equation}
\alpha_{\pm}=\frac{3}{2} \pm \sqrt{\frac{9}{4}-\frac{m^2 +\mu^2}{H^2} }\,
\end{equation}
and 
\begin{equation}
\bar\eta = |{\bf k}| \bar\tau
\end{equation}
where ${\bf k}$ is the wavevector associated to the shortest wavelenght mode that we consider in the bispectrum.
We will work in the limit $(m^2 +\mu^2)/H^2 \ll 1$, which implies that
\begin{equation}
\alpha_- \simeq \frac{m^2 +\mu^2}{3 H^2} \ll 1\,.
\end{equation}
We also assume that $\mu^2/(\mu^2+m^2) = \mathcal{O}(1)$ and
\begin{equation}
\frac{1-(-\bar\eta)^{\alpha_-}}{\alpha_-}\gg1\,.
\end{equation}
This last condition is required for the terms we keep in the power spectrum and bispectrum to be enhanced over those that we neglect.
Using the methods developed in \cite{an_non-gaussian_2018} we compute analytically the equal time correlation functions of the curvature perturbation at the end of inflation.

To compute correlation functions involving $\pi$ and $s$, we expand the quantum fields in terms of creation and annihilation operators. Due to the kinetic mixing term in the Lagrangian, the fields $\pi$ and $s$ share a pair of creation and annihilation operators with commutation relations,
\begin{equation}
[a^{(i)}({\bf p}), {a^{(j)}}^\dagger({\bf p'})]=(2 \pi)^3 \delta^{ij}\delta^{(3)}({\bf p - p'})\,.
\end{equation}
Introducing $\eta=|{\bf p}| \tau$ we write
\begin{equation}\label{eq:mode}
 \pi({\bf x},\tau)=\int \frac{d^3 p } {(2 \pi)^3}  \left( a^{(1)}({\bf p})  \pi_{|\bf p|}^{(1)}(\eta) e^{i{\bf p}\cdot {\bf x}}+a^{(2)}({\bf p}) \pi_{|\bf p|}^{(2)}(\eta)e^{i{\bf p} \cdot {\bf x}}  +{\rm h.c.}  \right)\,
\end{equation}
and
\begin{equation}\label{eq:mode2}
 s({\bf x},\tau)=\int \frac{d^3 p } {(2 \pi)^3} \left( a^{(1)}({\bf p})  {s}_{|\bf p|}^{(1)}(\eta) e^{i{\bf p}\cdot {\bf x}}+a^{(2)}({\bf p}) {s}_{|\bf p|}^{(2)}(\eta)e^{i{\bf p} \cdot {\bf x}}  +{\rm h.c.}  \right)\,.
\end{equation}
The mode functions $\pi_{|\bf p|}^{(i)}(\eta)$ and $s_{|\bf p|}^{(i)}(\eta)$ are determined by the equations of motion for the fields $\pi$ and  $ s $ and by the canonical commutation relations. For the calculation of the bispectrum when $\alpha_-$ is small it is the behaviour of these mode functions for $-\eta$ close to zero\footnote{Recall that $\eta$ is negative.} that is important \cite{an_non-gaussian_2018}. After rescaling the mode functions
\begin{eqnarray}
\label{rescaled mode functions}
\pi^{(i)}_{|\bf p|} (\eta) &=& \frac{H}{{|\bf p|}^{3/2}} \pi^{(i)} (\eta)\,,\\ s^{(i)}_{|\bf p|} (\eta)&=& \frac{H}{{|\bf p|}^{3/2}} s^{(i)} (\eta)\,
\end{eqnarray}
we can expand $\pi^{(i)}(\eta)$ and $s^{(i)}(\eta)$ in this region as
\begin{align}
\label{small mode function behavior}
\pi^{(i)}(\eta) &= a^{(i)}_{0}+ a^{(i)}_{-}(-\eta)^{\alpha_{-}} + a^{(i)}_{0,2}(-\eta)^{2}+a^{(i)}_{-,2}(-\eta)^{\alpha_{-}+2}+ a^{(i)}_{+}(-\eta)^{\alpha_{+}} + a^{(i)}_{3}(-\eta)^{3}  + \dots\,,\cr
s^{(i)}(\eta) &= b^{(i)}_{-}(-\eta)^{\alpha_{-}}+b^{(i)}_{0,2}(-\eta)^{2} + b^{(i)}_{-, 2}(-\eta)^{\alpha_{-}+2} + b^{(i)}_{+}(-\eta)^{\alpha_{+}}+b^{(i)}_{3}(-\eta)^{3} + \dots
\end{align}
where the ellipses represent terms with higher powers of $-\eta$ that we will not need. Using the equations of motion we get
\begin{align}
\label{btoas}
b^{(i)}_{0} = 0,\quad b^{(i)}_{-} = \frac{H a^{(i)}_{-} \alpha_{-}}{\mu},\quad b^{(i)}_{+} = \frac{H a^{(i)}_{+} \alpha_{+}}{\mu},\quad b^{(i)}_{3} = \frac{-3 H\mu}{m^{2}}a^{(i)}_{3}\,.
\end{align}
By matching this theory to an effective field theory in the small $-\eta$ limit \cite{an_non-gaussian_2018} it is possible to prove that
\begin{equation}
\sum_{i=1,2}|a_0^{(i)}|^2  = \sum_{i=1,2}|a_-^{(i)}|^2 =-\sum_{i=1,2}{\rm Re}[a_0^{(i)} a_-^{(i)*}]= \frac{9 \mu^2 H^2}{2(\mu^2 +m^2)^2}\,
\end{equation}
and by using the canonical commutation relations for the fields $\pi$ and $s$ we find 
 \begin{align}
     {\rm Im}[a_0^{(i)} b_3^{(i)*}]=\frac{\mu H}{ 2(\mu^2+m^2)},\quad {\rm Im}[a_-^{(i)*} b_+^{(i)*}]=-\frac{\mu H}{2(\mu^2+m^2)}\,.
 \end{align}
All other similar quantities are subleading in our calculations.
Using the above results the leading contribution to the power spectrum of the curvature perturbations in the limit of small $-\eta$ is
\begin{equation}
\label{eq:powerspectrum}
P_{\zeta}(\tau, |{\bf p}|)=\frac{9 H^6 \mu ^2 \left[1-(-\eta)^{\alpha_-}\right]^2}{2|{\bf p}|^3\dot\phi_0^2\left(\mu^2+m^2\right)^2}\,.
\end{equation}
This is needed to compute the impact of the change of coordinates from GC to CFC on the bispectrum (see the Appendix~\ref{app:A}).
\section{Bispectrum in Global Coordinates}\label{section:BGC}
In this section we work in GC and we compute the bispectrum for $\zeta$ in the squeezed limit at the end of inflation. 
Working to first order in the interactions and using the in-in formalism \cite{weinberg_quantum_2005} we have
\begin{align}
\label{in in}
\langle \zeta({\bar\tau, \bf x}_1)\zeta(\bar\tau, {\bf x}_2)\zeta(\bar\tau, {\bf x}_3) \rangle = i\int_{-\infty}^{\bar\tau}d\tau' \langle [H_{\rm int}(\tau') , \zeta(\bar\tau, {\bf x}_1)\zeta(\bar\tau, {\bf x}_2)\zeta(\bar\tau, {\bf x}_3) ] \rangle
\end{align}
where $H_{\rm int}$ denotes the interaction Hamiltonian in the interaction picture. In the squeezed limit we can drop the terms proportional to the spatial derivatives of $\pi$ from $\mathcal{L}_{\rm int}$ and the interaction Hamiltonian simplifies to
\begin{equation}
    H_{\rm int}(\tau)= \int d^3x  \frac{1}{(H\tau)^{2}\Lambda} (\partial_{\tau}\pi)^{2} s\,.
\end{equation}
Notice that $H_{\rm int}=\mathcal{L}_{\rm int}$ since we have a derivative interaction. Fourier transforming we find that the leading order contribution in $\alpha_-$ in the region of phase space that we are considering to the bispectrum in the squeezed limit is
\begin{equation}
B_{\zeta}^{(\rm GC)}({\bf q},{\bf k},{-\bf k}) \simeq -4 \left(\frac{H^7 \mu}{{\dot \phi}_0^4} \right) \frac{1}{|{\bf k}|^3 |{\bf q}|^3} \left(I_a+I_b+I_c \right)
\end{equation}
where
\begin{equation}
I_a = \int_{-1}^{\bar\eta}\frac{d \eta'}{ (-\eta')^2}{\rm Re}\left[ \pi^{(i)}\left(r \bar\eta \right) \dot{ \pi}^{(i)}\left(r \eta' \right)^*  \right]{\rm Re}\left[ \pi^{(j)} (\bar\eta)  
\dot{ \pi}^{(j)}\left(\eta' \right)^*  \right]{\rm Im}\left[ \pi^{(n)} (\bar\eta)    
{s}^{(n)}\left(\eta' \right)^*  \right],
\end{equation}

\begin{equation}
I_b = \int_{-1}^{\bar\eta}\frac{d \eta'}{ (-\eta')^2}{\rm Re}\left[ \pi^{(i)}\left(r \bar\eta \right) \dot{ \pi}^{(i)}\left(r \eta' \right)^*  \right]{\rm Im}\left[ \pi^{(j)} (\bar\eta)    
\dot{ \pi}^{(j)}\left(\eta' \right)^*  \right]{\rm Re}\left[ \pi^{(n)} (\bar\eta)    
{s}^{(n)}\left(\eta' \right)^*  \right],
\end{equation}
and
\begin{equation}
I_c = \int_{-1}^{\bar\eta}\frac{d \eta'}{ (-\eta')^2}{\rm Re}\left[ \pi^{(i)}\left( \bar\eta \right) \dot{ \pi}^{(i)}\left( \eta' \right)^*  \right]{\rm Im}\left[ \pi^{(j)} (\bar\eta)\dot{ \pi}^{(j)}\left(\eta' \right)^*  \right]{\rm Re}\left[ \pi^{(n)} \left(r \bar\eta \right)    
{s}^{(n)}\left(r \eta' \right)^*  \right]\,.
\end{equation}
In the above equations a dot indicates a derivative with respect to $\eta'$, the repeated mode function indices $i,j,n$ are summed over $1$~and~$2$ and we introduced the parameter $r \equiv{|{\bf q}|}/{|{\bf k}|}$. 
Most of the contribution to the integrals comes from the region $-\eta' \ll 1$ and to leading order in $(\mu^2 +m^2)/H^2$ we set the lower bound of the integrals to be $-1$.
Using the results of Section~\ref{sec:QSFI} we find that
\begin{equation}
{\rm Re}\left[ \pi^{(i)}\left(r \bar\eta \right) \dot{ \pi}^{(i)}\left(r \eta' \right)^*  \right]\simeq\frac{1}{ 2}\left(\frac{3 \mu^2}{\mu^2+m^2}\right) (-\eta')^{\alpha_--1} 
r^{\alpha_-} \left[1-\left(-r\bar\eta\right)^{\alpha_-} \right]\,,
\end{equation}
\begin{equation}
{\rm Re}\left[ \pi^{(i)}\left( \bar\eta \right) \dot{ \pi}^{(i)}\left( \eta' \right)^*  \right]\simeq\frac{1}{ 2}\left(\frac{3 \mu^2}{\mu^2+m^2}\right) (-\eta')^{\alpha_--1}
 \left[1-\left(-\bar\eta\right)^{\alpha_-} \right]\,,
\end{equation}
\begin{equation}
{\rm Im}\left[ \pi^{(i)}\left( \bar\eta \right) { s}^{(i)}\left( \eta' \right)^*  \right]\simeq\frac{1}{2}\left(\frac{H\mu  }{\mu^2+m^2}\right) \left[(-\eta')^3 - (-\eta')^{3-\alpha_- } (-\bar\eta)^{\alpha_-} \right]\,,
\end{equation}
\begin{equation}
{\rm Re}\left[ \pi^{(i)} \left(r \bar\eta \right)    
{s}^{(i)}\left(r \eta' \right)^*\right]\simeq\frac{3 H \mu  \left[(-r \bar\eta)^{\alpha_-}-1\right](-r\eta')^{\alpha_-}}{2\left(\mu^2+m^2\right)}\,,
\end{equation}
\begin{equation}
{\rm Re}\left[ \pi^{(i)} \left(\bar\eta \right)    
{s}^{(i)}\left(\eta' \right)^*\right]\simeq\frac{3 H \mu  \left[(-\bar\eta)^{\alpha_-}-1\right](-\eta')^{\alpha_-}}{2\left(\mu^2+m^2\right)}\,,
\end{equation}
\begin{equation}
{\rm Im}\left[ \pi^{(i)}\left(\bar\eta \right) \dot{ \pi}^{(i)}\left(\eta' \right)^* \right]\simeq\frac{\eta'^2\left[\mu^2\left(\frac{\eta'}{\bar\eta}\right)^{-\alpha_-}+m^2\right]}{2\left(\mu^2+m^2\right)}\,,
\end{equation}
and that ${\rm Im}\left[ \pi^{(i)} \left(r \bar\eta \right)    
{s}^{(i)}\left(r \eta' \right)^*\right]$ and ${\rm Im}\left[ \pi^{(i)}\left(r\bar\eta \right) \dot{ \pi}^{(i)}\left(r\eta' \right)^* \right]$ are suppressed in the squeezed limit.
Performing the $\eta'$ integration we have that to leading order in small quantities
\begin{equation}
I_a=\frac{27 H^3 \mu ^5 \left[(-\bar\eta )^{\alpha
   _-}-1\right]^3 r^{\alpha _-} \left[(-r\bar\eta)^{\alpha_-}-1\right]}{16 \left(\mu
   ^2+m^2\right)^4}
\end{equation}
and
\begin{equation}
  I_b=I_c= \frac{27 H^3\mu^3\left[(-\bar\eta)^{\alpha_-}-1\right]^2r^{\alpha_-} \left[(-r\bar\eta)^{\alpha_-}-1\right]\left[(-\bar\eta)^{\alpha_-}\left(2\mu^2+m^2\right)+m^2\right]}{16\left(\mu^2+m^2\right)^4}\,.
\end{equation}
This completes the calculation of the bispectrum in GC and we now turn to transform it to CFC.
\section{The Bispectrum in CFC}
We are now ready to compute the bispectrum in CFC by using Eq.~\eqref{eq:correctionB} to transform the result that we found in Section~\ref{section:BGC} for the bispectrum in GC. 
We have
\begin{equation}
B_\zeta({\bf q},{\bf k}, -{\bf k})=B^{(\rm GC)}_\zeta({\bf q},{\bf k}, -{\bf k})+ \Delta B_\zeta({\bf q},{\bf k}, -{\bf k})\,
\end{equation}
where
\begin{equation}
\Delta B_\zeta({\bf q},{\bf k}, -{\bf k})=P_{\zeta}(\bar\tau, |{\bf q}|)\frac{\partial}{\partial\log \bar\tau}P_{\zeta}(\bar\tau, |{\bf k}|)\,
\end{equation}
and to simplify the notation we dropped the superscript $\rm CFC$. 
Using Eq~\eqref{eq:powerspectrum} we get
\begin{equation}
\Delta B_\zeta({\bf q},{\bf k}, -{\bf k})\simeq\frac{27 H^{10} \mu^4 \left[(-\bar\eta)^{\alpha_-}-1\right](-\bar\eta)^{\alpha_-} \left[(-r\bar\eta)^{\alpha_-}-1\right]^2}{2 |{\bf k}|^3 |{\bf q}|^3\dot\phi_0^4\left(\mu^2+m^2\right)^3}\,.
\end{equation}
Even though for modes of cosmological interest $-\bar\eta =-|{\bf k}|\bar\tau \simeq e^{-60}$ \cite{baumann_tasi_2012}, $(-\bar\eta)^{\alpha_-}$ can still be of order unity, for example if $\alpha_- \sim 1/50$. In this case, the contribution to the bispectrum from the change of coordinates from GC to CFC, is comparable to the one that comes from the three point vertex in GC.

\begin{figure}[ht]
\includegraphics[scale=.8]{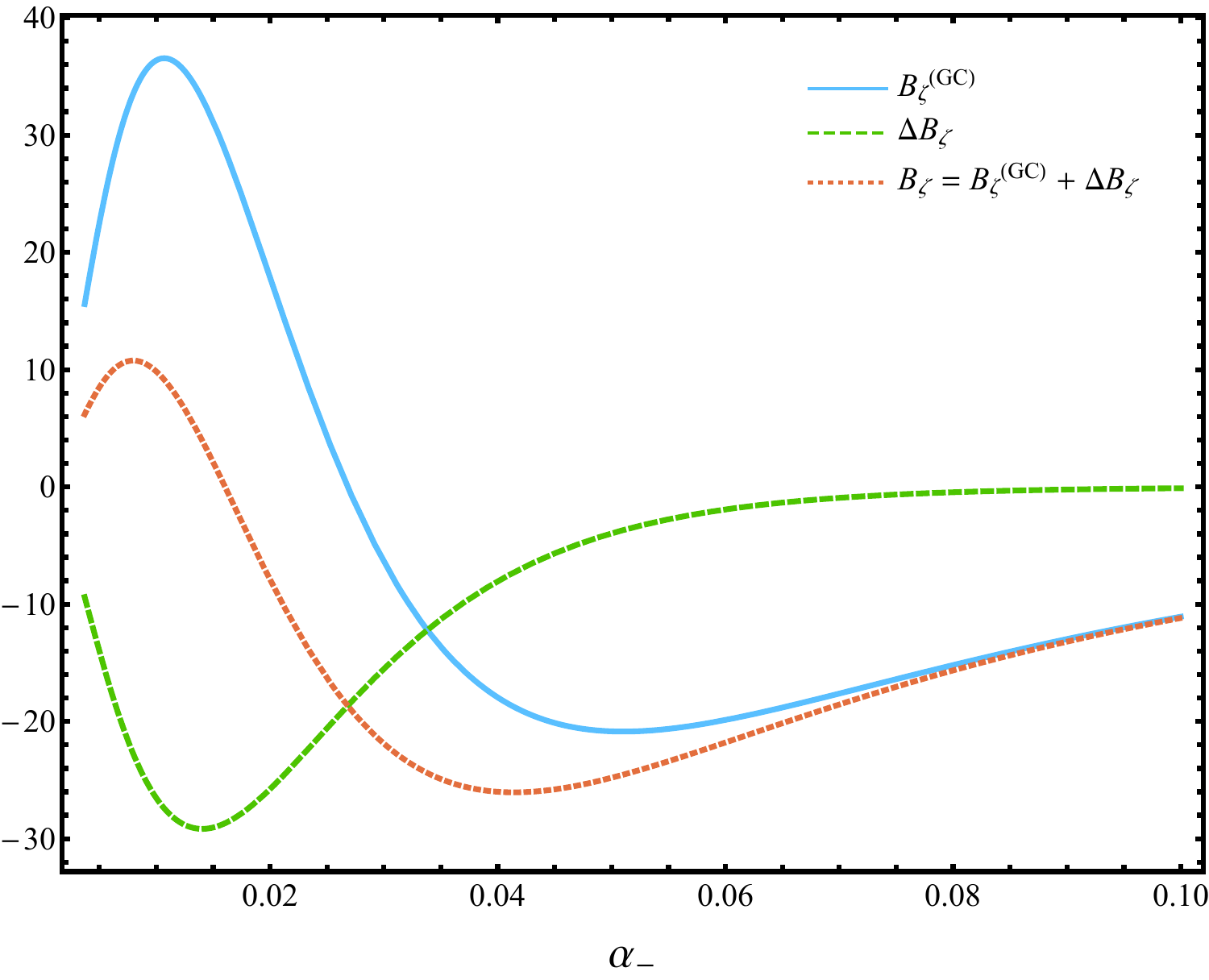}
\caption{Contributions to the Bispectrum for $m \ll \mu$, $-\bar\eta = e^{-60}$ and $r = 10^{-3}$. We consider values of $\alpha_-$ that go from $0.0037$ (corresponding to $(-\bar\eta)^{\alpha_-} = 0.8$) to $\alpha_- = 0.1$. The $y$-axis is in units of $H^8/(|{\bf k}|^3|{\bf q}|^3{\dot\phi}_0^4)$.}
\label{fig:B}
\centering
\end{figure}

In the limit $m \ll \mu$, we finally obtain
\begin{equation}
  B^{(\rm GC)}_\zeta({\bf q},{\bf k}, -{\bf k})=-\frac{1}{{|{\bf k}|}^3{|{\bf q}|}^3}  \frac{27 H^{10} \left[(-\bar\eta)^{{\alpha_-}}-1\right]^2 \left[5 (-\bar\eta )^{{\alpha_-}}-1\right] r^{{\alpha_-}}
   \left[(-r\bar\eta)^{{\alpha_-}}-1\right]}{4\mu^2{\dot\phi}_0 ^4}
\end{equation}
and
\begin{equation}
\Delta B_\zeta({\bf q},{\bf k}, -{\bf k})\simeq\frac{27 H^{10} \left[(-\bar\eta)^{\alpha_-}-1\right](-\bar\eta)^{\alpha_-} \left[(-r\bar\eta)^{\alpha_-}-1\right]^2}{2 |{\bf k}|^3 |{\bf q}|^3\dot\phi_0^4\mu^2}\,
\end{equation}
that are plotted in Fig.~\ref{fig:B}. Notice that the leading contribution to the bispectrum in GC vanishes for $(-\bar\eta)^{\alpha_-}=0.2$. At this point the part from the change of coordinates dominates the bispectrum.

\begin{figure}[ht]
\includegraphics[scale=.8]{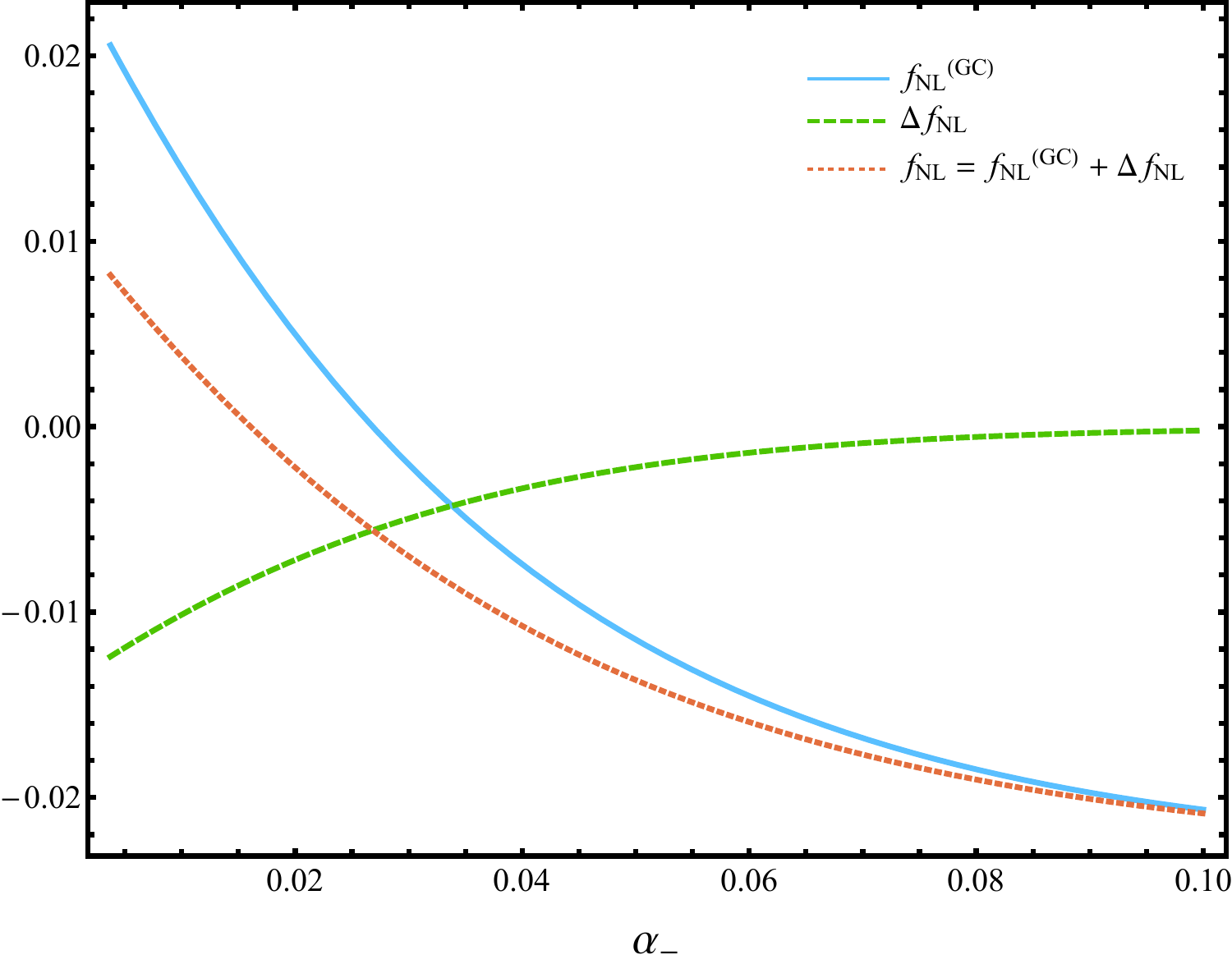}
\caption{Contributions to $f_{\rm NL}$ for $m \ll \mu$, $-\bar\eta =e^{-60}$ and $r = 10^{-3}$. We consider values of $\alpha_-$ that go from $0.0037$ (corresponding to $(-\bar\eta)^{\alpha_-} = 0.8$) to $\alpha_- = 0.1$.}
\label{fig:FNL}
\centering
\end{figure}

In Fig.~\ref{fig:FNL} we plot the local bispectrum $f_{\rm NL}$ in GC and CFC as a function of $\alpha_-$ where
\begin{eqnarray}
f_{\rm NL}^{(\rm GC)} = \frac{5}{12} \frac{B^{(\rm GC)}_\zeta({\bf q},{\bf k},-{\bf k})}{P_{\zeta}(\bar\tau, |{\bf k}|)P_{\zeta}(\bar\tau, |{\bf q}|)}\,,\\
\Delta f_{\rm NL} = \frac{5}{12} \frac{\Delta B_\zeta({\bf q},{\bf k},-{\bf k})}{P_{\zeta}(\bar\tau, |{\bf k}|)P_{\zeta}(\bar\tau, |{\bf q}|)}\,
\end{eqnarray}
and in the limit $m \ll \mu$ we have
\begin{eqnarray}
f_{\rm NL}^{(\rm GC)} &=&- \frac{5 \alpha_- r^{\alpha_-} \left[5 (-\bar\eta )^{\alpha_-}-1\right]}{12
   \left[(-r\bar\eta)^{\alpha_-}-1\right]}\,,\\
\Delta f_{\rm NL} &=& \frac{5 \alpha_- (-\bar\eta)^{\alpha_-}}{6 \left[(-\eta)^{\alpha_-}-1\right]}\,.
\end{eqnarray}
Both $B|{\bf k}|^3|{\bf q}|^3$ and $f_{\rm NL}$ depend very weekly on the value of $r$. This is because $r$ only enters in these quantities raised to the power $\alpha_-$.
\section{Concluding Remarks}
In this paper we considered QSFI where a dimension five operator couples the inflaton and the curvaton field. Working in the limit of small coupling and small curvaton mass we computed analytically the bispectrum in the squeezed limit in GC and in CFC. We found that transforming to CFC introduces a non-negligible correction to the result in GC. We also showed that $f_{\rm NL}$ can be either enhanced or suppressed by this effect, and in the region of parameter space that we considered $f_{\rm NL}\simeq 10^{-2}$. In this model $f_{\rm NL}$ is small and hence these non-Gaussianities could not be observed in the near future. However, this is an interesting example where the change of coordinates from GC to CFC can have an order one effect on the bispectrum.
\acknowledgments 
This material is based upon work supported by the U.S. Department of Energy, Office of Science, Office of High Energy Physics, under Award Number DE-SC0011632.
We are also grateful for the support provided by the Walter Burke Institute for Theoretical Physics. We thank Yi Wang and Zhong-Zhi Xianyu for useful comments.
\appendix
\section{Transformation of the Bispectrum to Conformal Fermi Coordinates}\label{app:A}
In this Appendix we rederive the coordinate transformation from GC to CFC and compute the bispectrum in CFC. Rather than taking the constructive approach of the previous literature we derive necessary and sufficient conditions that the coordinate transformation must satisfy.

The metric in GC is given by
\begin{equation}
g_{\mu \nu}(x)= a^2(\tau)\left[\eta_{\mu\nu} + h_{\mu\nu}(x) \right]\,
\end{equation}
and the metric scalar perturbations in $h_{\mu \nu}$ are expressed in terms of the curvature perturbation $\zeta$ as follows
\begin{eqnarray}
h_{00} &=& -2\frac{\partial_\tau \zeta}{\mathcal{H}}\,,\\
h_{0i}&=&-\partial_i\frac{\zeta}{\mathcal{H}}\,,\\
h_{ij}&=& 2\zeta \delta_{ij}\,,
\end{eqnarray}
where $\mathcal{H} \equiv \frac{1}{a}\frac{d a}{d\tau}$.
We split the metric perturbation as
\begin{equation}
h_{\mu\nu}(x) =h^L_{\mu\nu}(x)+h^S_{\mu\nu}(x)
\end{equation}
where $h^L_{\mu\nu}(k)\approx0$ for $k \gtrsim \Lambda$ and $h^S_{\mu\nu}(k)\approx0$ for $k \lesssim \Lambda$. Here $\Lambda$ is a cutoff that divides the modes into short and long.
In CFC with respect to the longest wavelength modes the metric has the form
\begin{equation}\label{eq:FermiMetric}
g^F_{\mu \nu}(x_F) = a^2(\tau_F)\left[\eta_{\mu \nu} +h^S_{\mu\nu}(x_F) +\mathcal{O}(x_F^i x_F^j) \right] \,
\end{equation}
where the terms $\mathcal{O}(x_F^i x_F^j)$ are made negligible by an appropriate choice of CFC. However this choice does not explicitly enter our analysis.

The coordinate transformation that takes us between these two frames can be expanded in $x_F^i$ as \cite{pajer_observed_2013,dai_conformal_2015,cabass_how_2017,bravo_vanishing_2018}
\begin{equation}\label{eq:coordtransf}
x^\mu(x_F) = x_F^\mu + \xi^\mu(\tau_F)+ A_{\;i}^\mu(\tau_F) x_F^i + B^\mu_{\;ij}(\tau_F)x_F^ix_F^j+ \mathcal{O}(x_F^i x_F^j x_F^k)
\end{equation}
where $\xi^\mu, A^\mu_{\;i}, B^\mu_{\; ij} = \mathcal{O}(h^L_{\mu\nu})$ and we neglect quantities $\mathcal{O}\left[(h^L_{\mu\nu})^2\right]$. Without loss of generality, we assume $B^\mu_{\; ij}(\tau_F) = B^\mu_{\; ji}(\tau_F)$.
The transformation law for the metric tensor
\begin{equation}
g^F_{\mu\nu}(x_F) = \frac{\partial x^\alpha}{\partial x^\mu_F}\frac{\partial x^\beta}{\partial x^\nu_F} g_{\alpha \beta}(x)\,
\end{equation}
gives ten differential equations for $\xi^\mu, A^\mu_{\;i}$ and $B^\mu_{\; ij}$ that need to be satisfied in terms of $h^L_{\mu\nu}$ in order for $g^F_{\mu\nu}(x_F)$ to have the form of Eq.~\eqref{eq:FermiMetric}. Requiring each differential equation to hold order by order in $x_F^i$ gives
\begin{eqnarray}\label{eq:DiffEqs}
(\partial_{\tau_F} + \mathcal{H}) \xi^0(\tau_F) &=&-\frac{\partial_{\tau_F} \zeta_L({\bf x}_F = {\bf 0}, \tau_F)}{\mathcal{H}(\tau_F)}\\
(\partial_{\tau_F} + \mathcal{H}) A_{\;i}^0(\tau_F) &=& -\frac{\partial_{\tau_F}\partial_i \zeta_L({\bf x}_F = {\bf 0}, \tau_F)}{\mathcal{H}(\tau_F)}\\
A^0_{\;i}(\tau_F)-\partial_{\tau_F}\xi_i(\tau_F) &=&-\partial_i\frac{\zeta_L({\bf x}_F = {\bf 0}, \tau_F)}{\mathcal{H}(\tau_F)}\\
2B^0_{\;ik}(\tau_F)-\partial_{\tau_F} A_{ik}(\tau_F)&=&-\frac{\partial_k\partial_i \zeta_L({\bf x}_F = {\bf 0}, \tau_F)}{\mathcal{H}(\tau_F)}\\
A_{ij}(\tau_F) + A_{ji}(\tau_F)+2\mathcal{H} \xi^0(\tau_F)\delta_{ij} &=& -2 \zeta_L({\bf x}_F = {\bf 0}, \tau_F)\delta_{i j}\label{eq:Aij}\\
B_{ijk}(\tau_F) + B_{jik}(\tau_F) +\frac{1}{2}\mathcal{H} A^0_{\; k}(\tau_F)\delta_{ij}&=& -\partial_k \zeta_L({\bf x}_F = {\bf 0}, \tau_F) \delta_{ij}\label{eq:DiffLast}
\end{eqnarray}
where the spatial indices were lowered using $\delta_{ij}$ and the quantities on the right hand side are the expressions in \emph{comoving} coordinates.
These are necessary \emph{and} sufficient conditions for Eq.~(\ref{eq:FermiMetric}) to hold.
With the coordinate transformation at hand we find how the connected three point function of $\zeta$ transforms in going from GC to CFC in the squeezed limit. Following \cite{bravo_vanishing_2018}, we have
\begin{eqnarray}
\langle \tilde\zeta^F(\tau_F,{\bf k}_1)\tilde\zeta^F(\tau_F,{\bf k}_2)\tilde\zeta^F(\tau_F,{\bf k}_3) \rangle=\int d^3{x}^F_1 d^3{x}^F_2 d^3{x}^F_3 e^{-i({\bf k}_1{\bf x}^F_1+{\bf k}_2{\bf x}^F_2+{\bf k}_3{\bf x}^F_3)} \langle \zeta^F(x_1^F)\zeta^F(x_2^F)\zeta^F(x_3^F) \rangle\nonumber
\end{eqnarray}
where $x^F_i \equiv (\tau_F,{\bf x}^F_i)$, $|{\bf k}_1| \ll |{\bf k}_2|,|{\bf k}_3|$.

Using spatial translational invariance we get
\begin{eqnarray}
&&\langle \tilde\zeta^F(\tau_F,{\bf k}_1)\tilde\zeta^F(\tau_F,{\bf k}_2)\tilde\zeta^F(\tau_F,{\bf k}_3) \rangle=(2\pi)^3 \delta({\bf k}_1+{\bf k}_2+{\bf k}_3)B^{(\rm CFC)}_\zeta({\bf k}_1, {\bf k}_2,{\bf k}_3) 
\end{eqnarray}
with 
\begin{eqnarray}\label{eq:interm}
B^{(\rm CFC)}_\zeta({\bf k}_1, {\bf k}_2,{\bf k}_3) &=& \int d^3{y}^F d^3{z}^F e^{-i\left[{\bf k}_1 {\bf y}^F+\left({\bf k}_3 + \frac{{\bf k}_1}{2}\right){\bf z}^F\right]} \left\langle \zeta^F\left(\tau_F,{\bf y}^F\right)\zeta^F\left(\tau_F,-\frac{{\bf z}^F}{2}\right)\zeta^F\left(\tau_F,\frac{{\bf z}^F}{2}\right) \right\rangle\nonumber\\*
&=&\int d^3{y}^F d^3{z}^F e^{-i\left[{\bf k}_1 {\bf y}^F+({\bf k}_3 + \frac{{\bf k}_1}{2}){\bf z}^F\right]}\langle0_L|\zeta_L(\tau_F,{\bf y}^F) \langle0_S
|\zeta_S(x_a)\zeta_S(x_b)|0_S\rangle|0_L\rangle
\end{eqnarray}
where $x_{a,b} = x_{a,b}(x_{a,b}^F)$ and $x^F_a = (\tau_F,-{\bf z}^F/2)$, $x^F_b = (\tau_F,{\bf z}^F/2)$\, and we assumed that $\zeta$ transforms as a scalar\footnote{We did not change the argument of $\zeta_L$ since it would have resulted in a disconnected piece that we discard.}. Moving forward we drop the designation $|0_S\rangle$ and write $\langle0_S
|\zeta_S(x_a)\zeta_S(x_b)|0_S\rangle\equiv \langle
\zeta_S(x_a)\zeta_S(x_b)\rangle$
in terms of $x^F_a$ and $x^F_b$ up to linear order in $\zeta_L$. Eq.~\eqref{eq:interm} implies that the contribution to the three point function is dominated by $|{\bf z}^F|\ll 1/|{\bf k}_3|$. Thus, using Eq.~(\ref{eq:coordtransf}) and working to linear order in the long mode, we find
\begin{eqnarray}
&&\langle 0_S| \zeta_S^F(x_a^F)\zeta_S^F(x_b^F)|0_S\rangle = \langle 0_S| \zeta_S(x_a^F)\zeta_S(x_b^F)|0_S\rangle+\Big[\xi^0(\tau_F) \partial_{\tau_F} + A^k_{\;i}(\tau_F) ({x_a^F}^i-{x_b^F}^i)\partial_k^{(a)}+\nonumber\\
&&+ \xi^i(\tau_F) (\partial_i^{(a)}+\partial_i^{(b)}) + \frac{1}{2}A^0_{\;i}(\tau_F)({x_a^F}^i+{x_b^F}^i)\partial_{\tau_F}\Big]\langle 0_S| \zeta_S(\tau_F,{\bf x}_a^F)\zeta_S(\tau_F,{\bf x}_b^F) |0_S\rangle
\end{eqnarray}
where the terms on the second line vanish because of translational invariance and because ${\bf x}^F_a = -{\bf x}^F_b$.
Using Eq.~\eqref{eq:Aij} we obtain
\begin{eqnarray}
 &&\langle0_S|\zeta_S^F(x_a^F)\zeta_S^F(x_b^F)|0_S\rangle = \langle0_S|\zeta_S(x_a^F)\zeta_S(x_b^F)|0_S\rangle+\\
&&+\left\{\xi^0(\tau_F) \partial_{\tau_F} -
\left[\mathcal{H}\xi^0(\tau_F)+\zeta_L({{\bf 0}}, \tau_F)\right]
{\bf z}^F\frac{\partial}{\partial {\bf z}^F}  \right\}\langle 0_S|\zeta_S(\tau_F, {\bf z}^F)\zeta_S(\tau_F, {\bf 0}) |0_S\rangle\,.\nonumber
\end{eqnarray}
Inserting this expression back in Eq.~\eqref{eq:interm} and using rotational invariance we get
\begin{eqnarray}
&&B^{(\rm CFC)}_\zeta({\bf k}_1, {\bf k}_2,{\bf k}_3)=
B^{(\rm GC)}_\zeta({\bf k}_1, {\bf k}_2,{\bf k}_3)- P_{\zeta}(\tau_F, |{\bf k}_1|)
\left(-3-\frac{\partial}{\partial\log |{\bf k}_3|}\right) P_{\zeta}(\tau_F, |{\bf k}_3|)\nonumber\\
&&+
 \langle\tilde\zeta^F_L(\tau_F,{\bf k}_1)\xi^0(\tau_F)\rangle\left[\partial_{\tau_F} -\mathcal{H}\left(-3-\frac{\partial}{\partial\log |{\bf k}_3|}\right) \right]P_{\zeta}(\tau_F, |{\bf k}_3|)
\end{eqnarray}
where
\begin{equation}
\langle \tilde\zeta(\tau_F,{\bf k}_1)\tilde\zeta(\tau_F,{\bf k}_2)\tilde\zeta(\tau_F,{\bf k}_3)\rangle=(2\pi)^3 \delta({\bf k}_1+{\bf k}_2+{\bf k}_3) B^{(\rm GC)}_\zeta({\bf k}_1, {\bf k}_2,{\bf k}_3)
\end{equation}
and we used that $|{\bf k}_1| \ll |{\bf k}_3|$. In the limit in which scale invariance is preserved Eq.~\eqref{eq:scaleinvaraince} and the fact that $\mathcal{H} \tau_F=-1$ imply that the final result does not depend on the integration constants of Eqs.~\eqref{eq:DiffEqs}-\eqref{eq:DiffLast}. We finally obtain
\begin{eqnarray}\label{eq:correctionB}
B^{(\rm CFC)}_\zeta({\bf k}_1, {\bf k}_2,{\bf k}_3)=
B^{(\rm GC)}_\zeta({\bf k}_1, {\bf k}_2,{\bf k}_3)+ P_{\zeta}(\tau_F, |{\bf k}_1|)\frac{\partial}{\partial\log \tau_F}P_{\zeta}(\tau_F, |{\bf k}_3|)\,.
\end{eqnarray}
This expression coincides with the one in \cite{bravo_vanishing_2018} for scale invariant models of inflation.

\bibliographystyle{ieeetr}
\bibliography{references.bib}

\end{document}